# Status of CMS and preparations for first physics

A. H. Ball (for the CMS collaboration)
*PH Department, CERN, Geneva, CH1211 Geneva 23, Switzerland*

The status of the CMS experiment is described. After a brief review of the detector design and a short overview of the first 5 years of assembly, the focus of this presentation will be the parallel activities of completing and commissioning over the last 2 years and the readiness of CMS for the exciting prospect of first LHC operation.

## 1. INTRODUCTION

Based on simulation of benchmark processes, such as the decays of Z' and Higgs bosons and of supersymmetric particles, produced in pp collisions at the Large Hadron Collider (LHC), the Compact Muon Solenoid (CMS)[1] is designed as a general purpose physics detector for LHC, having 4 main qualities:

A. Efficient, hermetic muon triggering & identification with low contamination and good momentum resolution, specifically di-muon mass resolution < 1% at 100GeV/$c^2$, charge determination for muons with momentum ~ 1 TeV/c and $\Delta p_T/p_T$ ~ 5%.

B. A powerful central tracking system with good reconstruction of secondary vertices to detect the decays of long-lived b quarks and tau leptons.

C. Highly granular, hermetic electromagnetic calorimetry with good energy resolution, ~ 0.5% at $E_T$ ~ 50 GeV and di-photon mass resolution < 1% at 100 GeV/$c^2$.

D. An hermetic overall calorimetry system, giving good resolution for detecting and measuring "missing" $E_T$ and reconstructing the mass of jet-pairs.

The first requirement drives the design of the magnet and thus the mechanical structures supporting the detector systems. The second and third are difficult in the LHC environment and need specially developed technologies. The fourth, combined with aspects of the first, pose severe integration challenges.

## 2. DESIGN AND CONSTRUCTION

### 2.1 General Layout

The engineering solution to the requirements listed above is illustrated in Figure 1. It is founded on a very large 4 Tesla superconducting solenoid [1,2], 6m in diameter and 13m long, the largest practicably constructible, and its return yoke. This leads to a design which is comparatively compact (14 x 22 m), yet massive (12,500 t). It provides sufficient $BL^2$ to achieve the required muon momentum resolution without excessive demands on the resolution and alignment of the muon detection stations. It also allows the barrel tracking and calorimetry to be contained within the solenoid. As a consequence the solenoid can be thick without affecting performance.

The return yoke is accurately constructed and instrumented with 4 layers of muon detection [1,3], each designed to allow reconstruction of track position and trajectory with plenty of redundancy. Drift tube chambers (DT) are used in the central region (|η|< 1.2), whilst Cathode Strip Chambers (CSC) are used in the more challenging forward region (1.2<|η|< 2.4), where the muon rate, the background and the fringe magnetic field in the detectors are all higher. In both

regions, layers of Resistive Plate Chambers (RPC), with excellent time resolution, but coarser position measurement, provide trigger information complementing that given by the DT's and CSC's. Within the solenoid vacuum tank, the beam pipe, tracker, electromagnetic calorimeter (ECAL), and hadron calorimeter (HCAL) are arranged in radial sequence. The central charged particle tracking system [1,4] is all solid-state on an unprecedented scale. The main part of its 2.6m diameter, 5.8 m long cylindrical volume comprises 10 layers of silicon microstrips (200m$^2$ in area, approximately 10M channels), while the inner layers at 4, 7 and 11 cm radius are instrumented by silicon pixels (66M channels), which provide space points used as seeds for the recognition and reconstruction of tracks. The crystal electromagnetic calorimeter [1,5] uses a matrix of nearly 70,000 lead tungstate (PbWO$_4$) crystals, arranged as a barrel and two endcaps. The short radiation length (8.9mm) allows the barrel section to achieve the required performance using a minimum of the limited radial space within the solenoid vacuum tank. The fast scintillation light is read out by avalanche photodiodes in the barrel, whilst vacuum phototriodes are used in the endcaps. A preshower detector [1,5], using 2 layers of silicon microstrips and lead pre-radiator is installed in front of each endcap, principally to help identify neutral pions. The sampling hadron calorimeter [1,6] is constructed to cover the region $|\eta|$ <5 as hermetically as possible. In the barrel (1.3<$|\eta|$) and endcaps (1.3<$|\eta|$< 3.0), it is made of interleaved layers of brass and scintillator, the latter being read out via embedded wavelength shifting fibres, clear fibres and hybrid photo-diodes. In the forward region, (3.0<$|\eta|$< 5.0), the absorber material is iron, instrumented with a dense matrix of quartz fibres, whose Cerenkov emissions are read out by photomultiplier tubes.

The compact design lent itself to modular construction (Figure 1) allowing pre-assembly in 15 large modules [2] in a surface assembly building, independent of the difficult civil engineering excavation for the underground LHC collision hall 5. Provision was made for surface testing of the solenoid and yoke closure before lowering the largely pre-assembled elements, varying in weight from 350t to 2000t, 100m into the more restricted underground environment.

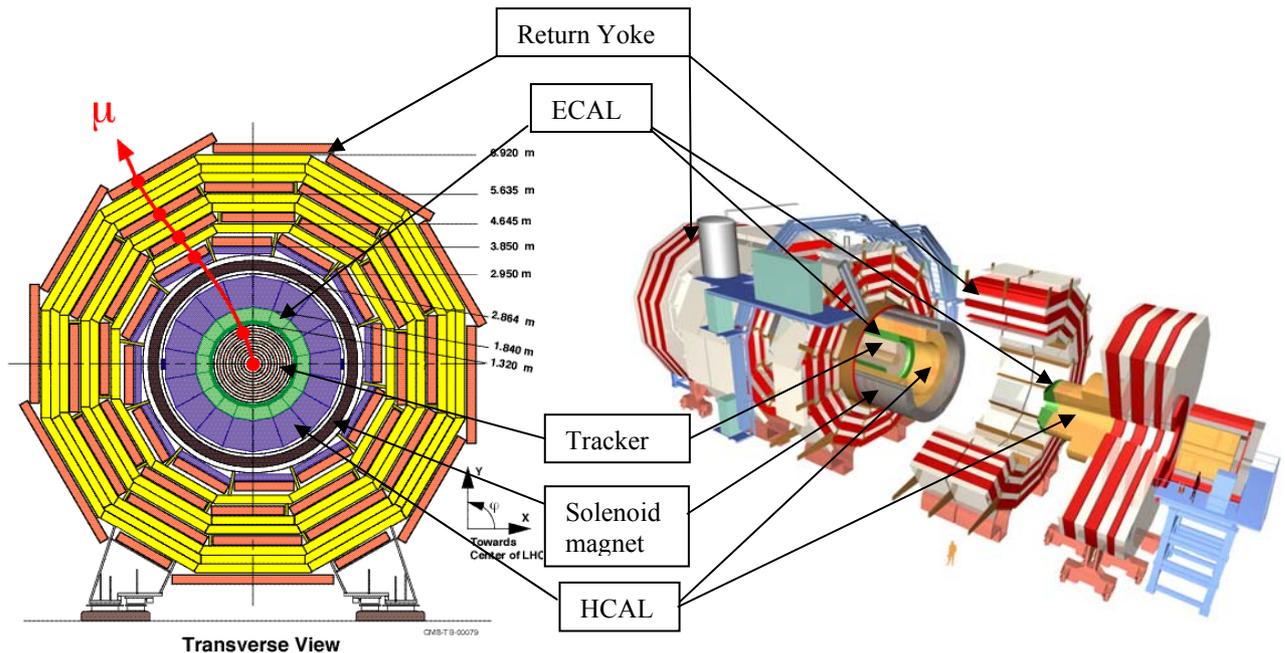

Figure 1: Transverse view and modular construction of the CMS experiment

## 2.2 Surface Assembly, Pre-Commissioning and Lowering

The main elements were assembled in the SX5 assembly hall at Cessy, France, over a seven year period, beginning in 2000, starting with the magnet and return yoke and progressing to the hadron calorimeter and the muon stations within the yoke. The "bare" experimental cavern was delivered in 2004, although a further 2 years were needed to equip it with infrastructure and pre-install the cable chains which would later connect the major elements to the neighbouring service cavern. In 2006, the yoke was closed and the magnet fully tested along with parts of all subsystems and prototype trigger [1,7] and data acquisition systems [1,8] in the first major proof of successful system integration. This activity required a full rehearsal of calorimeter and tracker installation, as well as the delicate closing sequence. During this period the magnetic field inside the solenoid was measured, leading to an operational field map with a precision better than 2mT. The operation together of the solenoid and representative parts of each subsystem in a "cosmic challenge" [11] was a crucial step in establishing mutual compatibility of systems and marked the beginning of a fully parallel effort in completion and commissioning which has continued for the last 2 years.

As assembly continued, the 15 elements (varying in weight between 350t and 200t) were lowered sequentially into the underground cavern UXC55 between November 2006 and January 2008 using a specially built gantry-crane employing the strand-jacking technique. Surface testing of the remaining elements with cosmic rays continued throughout this period, using facilities set up for the "cosmic challenge".

## 2.3 Underground Assembly

Elements were connected to their cable chains immediately after lowering so that re-commissioning could begin. The modular structure, easily moveable on air-pads, lends itself to a variety of configurations where several parts of the detection system, moveable without disconnection, can be worked on simultaneously. Once the 2000t central section was underground and aligned with the nominal beam axis to $\leq 1$ mm precision, detectors inside the solenoid were installed from outside in, starting with the two pre-assembled 500t barrels of the hadron calorimeter. Pre-installation of service pipes, cables and fibres on the central wheel could then start, in layers with the uppermost services serving the innermost detectors. This activity was one of the main schedule drivers during 2007, with more than 50,000 man-hours invested. Concurrently the 36 supermodules of the electromagnetic calorimeter were fitted inside the hadron calorimeter and connected to services at local racks and to patch panels on the solenoid vacuum tank. Pre-installed services included those of the tracker, terminated at patch panels inside the vacuum tank, so that, when the silicon strip tracker arrived after surface pre-commissioning, reconnection was relatively fast, using short pipes, cables and fibres. With the strip tracker in place, the beampipe could be installed. Configuring the experiment for bakeout of the pipe involved the first displacement of elements, particularly the 1500t first endcap disk, over many metres with the beampipe in place and involving radial clearances as little as 30mm. The successful completion of these moves was first proof of a key element in the CMS maintenance strategy. The long process of removing, replacing and re-baking the beampipe is only necessary in case of a fault with the pipe or the need to re-activate its "non-evaporable-getter" (NEG) inner coating. For maintenance access to CMS, the pipe is re-filled to 1atm pressure with pure Neon gas.

With the bakeout complete and the endcap open again, there started one of the most exciting and challenging few months in the 8 year assembly history of CMS, as the access tooling was reconfigured to install both the pixel tracker and the electromagnetic calorimeter endcaps in parallel, The pixel barrel was signed off in late July and the endcap just 24 hours before the material in this paper was presented. Meanwhile, the dee-shaped elements of the electromagnetic endcap were being mounted within 24 hours of arrival from the assembly facility. The last of the four was successfully installed on the last day of July and cabling was in full swing during the ICHEP meeting.

The solenoid magnet, integrated into the central barrel wheel, was lowered in February 2007. Lowering of the cold box followed shortly thereafter and reconnection and re-commissioning of the cryogenic, electrical and control systems began. At the time of this conference, the coil was stable at its operating temperature and all mechanical, electrical and control system tests were complete.

## 3. COMMISSIONING AND PERFORMANCE

In parallel, and generally parasitic to the continuing lowering and installation work, commissioning of the detector as an integrated system continued. Just 6 months after the first element were lowered, the first cosmic triggers were established underground. In September 2007, just one year after the surface integration test, the same level of complexity was reached in the underground commissioning, although it took a further 10 months to have representation from all of the major subsystems, in line with the continued detector installation process. This evolution is represented in Figure 2.

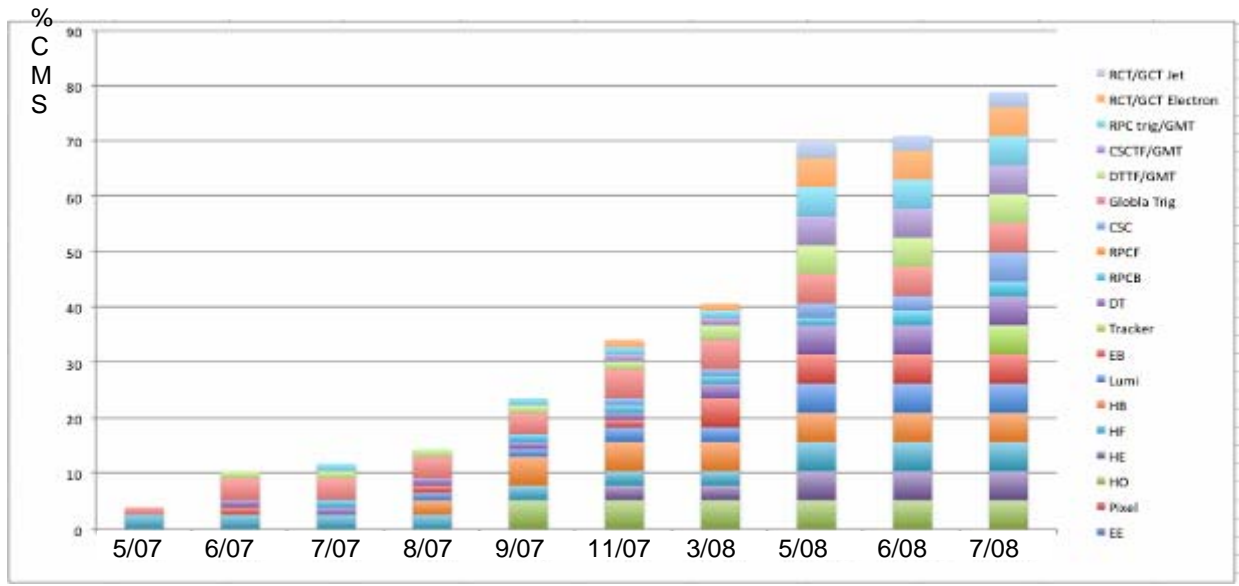

Figure 2: Percentage of CMS systems participating in regular, global, cosmic data-taking exercises
A total of 19 separate, equally weighted DAQ or readout elements contribute, according to % completeness

The following section gives a flavour of the status today and the performance currently being obtained.

The drift tube chambers of the barrel muon detection system have been fully commissioned with cosmic rays and 99.8% of the 17,000 channels are operating. This is particularly encouraging because the barrel yoke wheels were fully closed and locked in April 2007, with the schedule allowing no further opportunities for maintenance access.

Data acquisition [1,8] and trigger[1,7] are fully integrated and the system routinely and reliably provides a cosmic ray trigger, with a stable rate of about 200Hz, for local tests of subdetectors and global tests of the integrated apparatus.

Some difficulties were encountered in starting the gas re-circulation and filtering system for the resistive plate trigger chambers (RPC's) of the barrel muon system, but it is now functioning well. 4 of the 5 barrel wheels are working with 97% of channels active and work on the 5'th wheel progressing fast. Trigger [1,7] and Data acquisition [1,8] are fully integrated and the noise, resolution and efficiency are consistent with expectations. Work on the endcaps disks has been given lower priority, with an objective of having one endcap fully functional for first LHC beams.

The cathode strip chambers of the endcap muon system are fully functional with 97% of channels in operation, trigger [1,7] and data acquisition [1,8] fully integrated and all faults identifiable and reparable once access is available. This was the first CMS system operating on the surface, the first to fully integrate with central systems and it has mature calibration and diagnostic techniques in place. For instance, the individual timing adjustment for each of the 468 chambers can be derived from just two days of cosmic ray data.

The resolution and energy calibration of the hadron calorimeter was fully explored in test beams [9], whilst after surface pre-assembly, a very important pre-calibration of the detector at the 4-5% level was performed using moving wire sources. Underground, all (100%) of channels are operating and the overall response to cosmic rays triggered by the muon system matches that from test beam, so that the initial energy calibration can be established. In fact this cosmic ray calibration correlates in detail in $\eta$ vs $\phi$ with the surface source calibration. Trigger [1,7] and data acquisition [1,8] are fully integrated.

All modules of the barrel electromagnetic calorimeter were calibrated [9] with cosmic rays on the surface, establishing a very good initial inter-calibration of about 1.5%. A quarter of the 36 supermodules were also extensively beam-calibrated [9] so that the response vs energy is well understood. Due to the encouragingly low noise, re-commissioning was possible using cosmic rays, which leave a typical 250MeV in the calorimeter. The occupancy of 3x3 crystal matrices around a 70MeV seed signal is shown in Figure 3 for the whole barrel calorimeter, and shows that it is fully active with just 0.14% masked channels. Trigger [1,7] and data acquisition [1,8] are fully integrated.

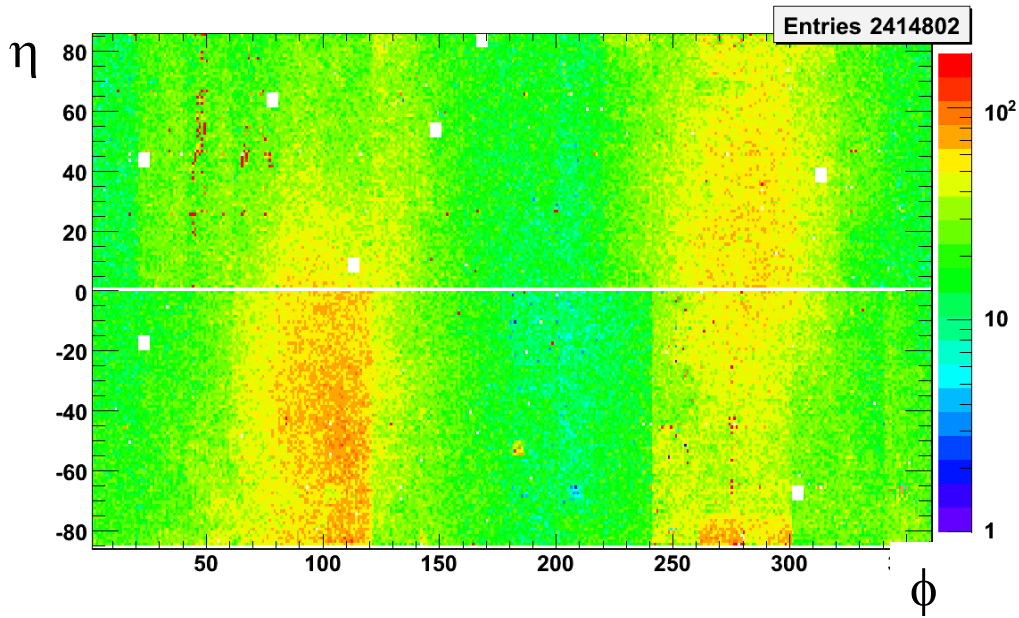

Figure 3: Electromagnetic calorimeter occupancy map: 3 x 3 matrix around >70 MeV seed

The silicon strip tracker engaged in an extensive surface commissioning programme using 20% of the detector. The noise performance (3 per mil) and the layer efficiency (99.8%) were found to be excellent. The signal to noise ratio (> 25:1) is well within specification and the tracking performance was outstanding over a large temperature range. To date the same encouraging performance is being found in the re-commissioning of the complete detector underground. This underground commissioning was delayed 2-3 months due to failure and subsequent repair of the cooling plant. The data acquisition is fully integrated, and the alignment algorithms are working well.

As a summary of the status, Figure 4 is an event picture taken from the July 2008 global cosmic run. Such a picture is a good benchmark of detector readiness, since it depends on many different parts of the system being ready from detector signals through trigger [1,7] and data acquisition [1,8] through the software [1,9] for dataflow and reconstruction.

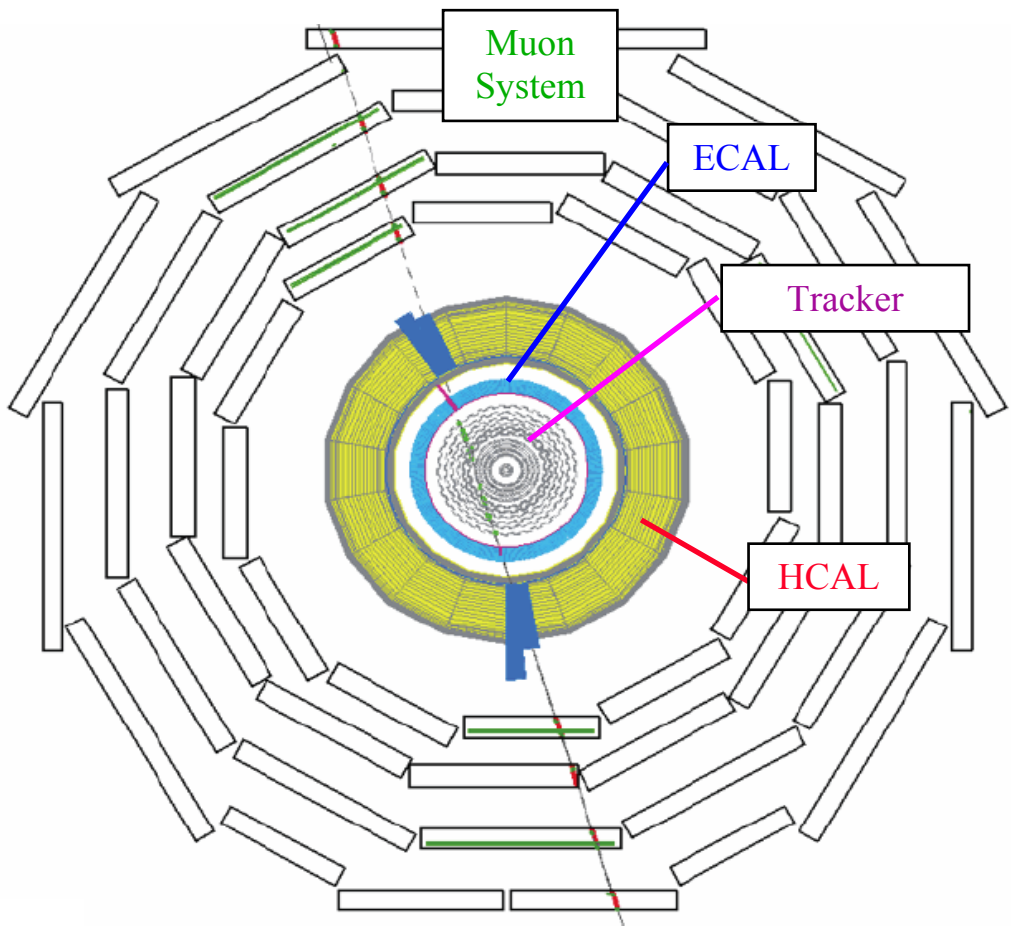

Figure 4: Cosmic Ray, triggered by barrel muon detector and seen by all CMS subsystems, reconstructed by global track-fitting algorithm

## 4. COMPLETION

The configuration in early August 2008, after cabling and commissioning of the recently installed pixel tracker and endcap calorimeter, corresponds to a typical future shutdown maintenance scenario, with a standard set of steps to prepare CMS for beam operation.

After removal of tooling and access equipment, the endcap yokes are closed by a 10m movement along the beampipe. Thereafter the forward calorimeters are brought out of their shielded garages and raised progressively to beam height on a series of stacking platforms. Finally the forward shielding structure is closed around the beamline elements between these calorimeters and the first machine quadrupole. Depending on the LHC programme planned, the very forward calorimeter, CASTOR, and the forward tracking systems of the TOTEM collaboration may also be installed, although only prototype elements are available for 2008. On closure of the yoke and shielding, the magnet is re-commissioned to its operating field of 3.8T. CMS is then ready for beam.

## 5. PREPARATION FOR FIRST BEAMS

### 5.1 Programme for Initial Beam Operation

The programme for using the period from first beams up to 10pb$^{-1}$ starts with 5 basic steps:

i) Commissioning the beam radiation monitoring system including abort
ii) Tuning operating procedures for beam operation
iii) Establishing the (lack of) effect of the CMS solenoid field on LHC beams
iv) Synchronizing detectors using beam timing
v) Commissioning the beam trigger and starting "physics commissioning"

Physics commissioning will begin with intensive alignment and calibration activity using beam-halo events and a large minimum-bias event sample. Once the basics are understood, jet and lepton rates will be measured and the identification of standard known signatures such as W, Z and top will begin. Even during this period a first look for possible extraordinary signatures will be made.

### 5.2 Tests of workflows and computing infrastructure

The important first step of this programme has been rehearsed as fully as possible during 2008 in two interlinked exercises. The first was the Computing, Software and Analysis exercise CSA08, which tested the offline workflows needed using simulated data. Samples of minimum-bias, noise, QCD jets, cosmic, $J/\psi$, Z, events were generated. focussing on the 1, and 10 pb$^{-1}$ total sample scenarios and simulating the most likely starting conditions for the CMS detector. Thereafter, data selection & reduction algorithms were applied, giving an opportunity to test alignment & calibration processes, latency and reprocessing with derived constants. An example output is shown in Figure 5, where the p$_T$ resolution of the silicon strip Tracker, starting from the survey data and startup geometry, is shown to progressively reach $\sigma \approx 2.2$GeV using muon tracks from 10pb$^{-1}$ of simulated data.

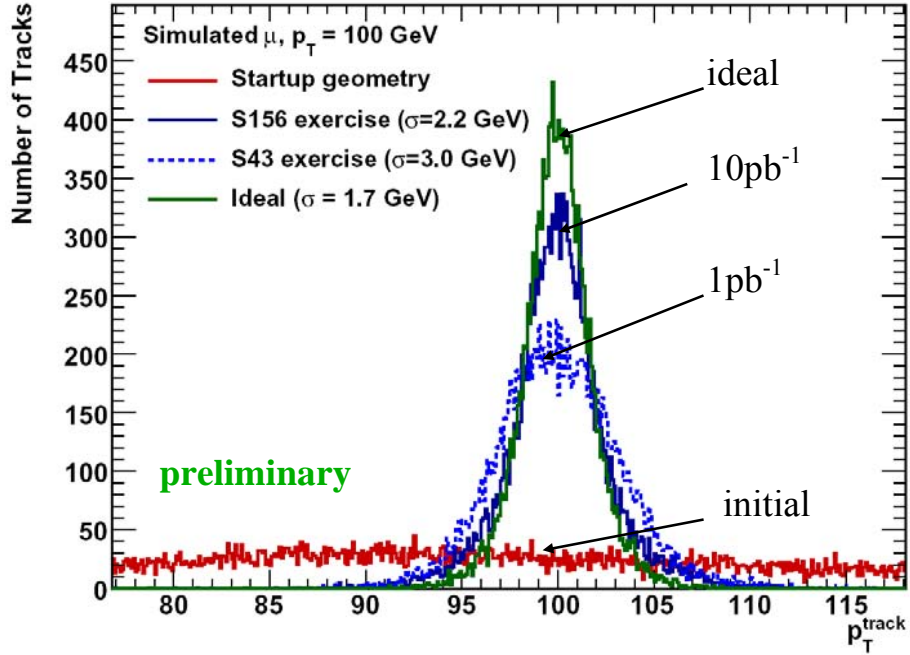

Figure 5: Tracker $p_T$ resolution at 100 GeV for various data samples

Parallel, quasi-real-time physics analysis of a few selected topics was also done, concentrating on the inclusive charged particle spectrum, measurement of the underlying event, measurements of the inclusive jet spectrum and di-jet angular distribution and finally the observation of dilepton peaks (J/psi, Upsilon, Z…). This exercise demonstrated CMS readiness for real data from LHC and provided the CMS workflow input to the CMS Common Computing Readiness Challenge CCRC 08 conducted in May. This was a stress test of computing infrastructure for data-processing at LHC (simultaneously with other experiments) and involved all aspects of data-flow and data storage thorough to the TIER 2 grid centres. The infrastructure performed well and CMS contributed the majority of the throughput throughout the exercise.

## 6. CONCLUSION

Construction of the CMS experiment is almost completed. Commissioning work already carried out gives confidence that CMS detectors will operate with the expected performance [10]. Integrated operation of subdetectors and central systems using cosmic triggers is routine with near-final complexity and functionality. Challenges conducted around the clock at 100% of 2008 load show that Computing, Software & Analysis tools are ready for early data. Preparations for the rapid extraction of physics are being made. CMS will be closed with magnetic field on, taking cosmic data, when LHC beams finally circulate.

### Acknowledgments

I would like to thank the organizers for their invitation, their hospitality and their efficiency in making ICHEP08 a pleasurable and rewarding event.

The construction of CMS has required an unprecedented effort by a very large, diverse, international team, supported by their various funding agencies, over a period of more than a decade. It is a privilege and a huge responsibility to be asked to represent their work to the world. Fortunately, my task was to narrate the latest chapter of a story of outstanding success, at times achieved against the odds, and I would like to thank the whole collaboration for their part in providing such material.

## References

[1] CMS Collaboration, "The CMS experiment at the CERN LHC", S. Chatrchyan et al,
    J. Instrum. 3 (2008) S08004.
[2] CMS Collaboration, "Magnet Technical Design Report", CERN/LHCC 1997-010 (1997).
[3] CMS Collaboration, "Muon Technical Design Report", CERN/LHCC 1997-032 (1997).
[4] CMS Collaboration, "Tracker Technical Design Report", CERN/LHCC 1997-006 (1998)
   Modified by Addendum CERN/LHCC 2000-16 (2000)
[5] CMS Collaboration, "ECAL Technical Design Report", CERN/LHCC 1997-033 (1997).
[6] CMS Collaboration, "HCAL Technical Design Report", CERN/LHCC 1997-031 (1997).
[7] CMS Collaboration, "Level 1 Trigger Technical Design Report", CERN/LHCC 2000-38 (2000).
[8] CMS Collaboration, "Data Acquisition and High Level Trigger Technical Design Report",
    CERN/LHCC 2002-26 (2002).
[9] CMS Collaboration, "Physics Technical Design Report, Vol1; Detector performance and software",
   CERN/LHCC 2006-001 (2006).
[10] CMS Collaboration, "Physics Technical Design Report, Vol2; "Physics performance",
    CERN/LHCC 2006-021 (2006).
[11] T. Christiansen, "The CMS Magnet Test and Cosmic Challenge", IEEE Nuclear Science Symposium
   and Medical Imaging Conference, San Diego, California, Oct 2006.